\documentclass{cernyrep}
\usepackage{texnames}
\usepackage[T1]{fontenc}

\usepackage{varwidth}
\usepackage{xcolor}

\usepackage{amsbsy}
\usepackage{dcolumn}
\newcolumntype{d}{D{.}{.}{2.6}}
\newcolumntype{s}{D{.}{.}{2.1}}

\begin{document}

\title{Laser-driven Plasma Wakefield: Propagation Effects}

\author{B. Cros}

\institute{Laboratoire de Physique des Gaz et des Plasmas, CNRS-Universit\'{e} Paris Sud, Orsay, France}

\maketitle 

\begin{abstract}
In the frame of laser-driven wakefield acceleration, the main characteristics of  laser propagation and plasma wave excitation are described,  with an em\-phasis on the role of propagation distance for electron acceleration.
To optimize inter\-action length and maximize energy gain, operation at low plasma density is the most promising regime for achieving ultra-relativistic energies. Among the possible methods of extending propagation length at low plasma density, laser guiding by grazing incidence reflection at the wall of dielectric capillary tubes has several assets. The properties of laser guiding and  the measurement of plasma waves over long distances are presented.
\end{abstract}

\section{Introduction}

Propagation effects play an important role in laser-driven plasma accelerators \cite{tajima79}. The mechanism of plasma wave excitation relies on non-linear effects \cite{chen83}, linked to the presence of  background  electrons in the medium, driven by short and intense laser pulses \cite{esar09}. Plasma waves result from the action of the pondero\-motive force, which is proportional to the gradient of laser energy. This force expels electrons from the regions of higher intensity. Depending on the laser intensity, the density perturbation, or wakefield, remaining behind the laser pulse may be a periodic plasma wave, \eg a sinusoidal oscillation \cite{andreev92},  or cavities void of electrons, also known as bubbles \cite{pukhov02}. Laser and plasma parameters need to be carefully selected to take advantage of non-linear effects.

The high intensity value, typically in the range $10^{17}$ to $10^{19} \, $W/cm$^2$, required to drive a wakefield by laser is usually achieved by focusing the laser beam, within a small volume around the focus position, over a length shorter than or of the order of $1 \Umm$.  The longitudinal fields associated with plasma waves driven by laser can be as large as $100 \UV[G]/\UmZ$, leading to accelerated electrons with energies of the order of $100\UMeV$ over a typical scale length of $1\Umm$ \cite{mangles04, geddes04, faure04}. Acceleration of electrons to ultra-high energies requires that a high acceleration gradient be maintained over a longer distance \cite{leemans14}.

Increasing the acceleration distance is one of the current challenges of laser plasma accelerators. Ultra-intense laser beam interaction with matter gives rise to various kinds of non-linear effect \cite{esar09}, which usually increase with propagation distance. Some intrinsic limitations of laser-driven wakefields, such as dephasing  of electrons during the acceleration process as they overrun the accelerating phase of the field, or depletion of the laser beam over the propagation distance, can be less severe at low plasma densities. Lowering the plasma density contributes to a reduction in non-linear effects and an increase in acceleration distance, provided the laser intensity can be maintained at the required level over the whole distance. This can be achieved by guiding the laser beam externally, by either a preformed plasma structure or a capillary tube.

The first part of this paper outlines some of the main characteristics of laser-driven plasma wakefields, useful for describing laser propagation and plasma wave excitation. The main parameters gov\-ern\-ing electron acceleration are also described, with an emphasis on the role of propagation distance.
Among the different methods of optimizing  interaction length to maximize energy gain, operation at low plasma density is the most promising regime towards achieving ultra-relativistic energies. Laser guiding by grazing incidence reflection at the wall of dielectric capillary tubes is described in the second part of this paper, from guiding properties to the measurement of plasma waves over long distances.

\section{Laser plasma acceleration characteristics}

\subsection{Laser propagation in a vacuum}

The electromagnetic field of a laser is described
by Maxwell's equations \cite{saleh07}. The laser pulse can be modelled by Gaussian functions in space and time as a good approximation of experimental profiles. The electric field $\textbf{E}$ of  a bi-Gaussian laser beam, propagating in vacuum along the $ z$ axis, is given by:
\begin{equation}
\begin{split}
\mathbf{E}(r,z,t)=& E_\mathrm{L}\frac{w_0}{w(z)}\exp\left[-\frac{r^2}{w^2(z)}\right]\exp\left[-2\ln(2)\frac{(z-ct)^2}{c^2\tau_0^2}\right] \\
& \times\mathfrak{R}\left\{\exp\left[\mathrm{i}\omega_0t-\mathrm{i}k_0z-\mathrm{i}k_0\frac{r^2}{2R(z)}+\mathrm{i}\psi_\mathrm{g}(z)\right]\mathbf{e}_{\bot} \right\}\ ,
\end{split}
\label{eq:Efield}
\end{equation}
where $E_\mathrm{L}$ is the amplitude of the electric field, $w_0$ is the waist or smallest laser transverse size in the focal plane ($z=0$), $c$ denotes the speed of light in a vacuum, $\tau_0$ is the full width at half maximum of the pulse duration, and $k_0=2\pi/\lambda_0$ and $\omega_0=ck_0$ represent, respectively, the wavenumber and angular frequency of a laser beam with wavelength $\lambda_0$. The unit vector $\mathbf{e}_{\bot}$ indicates the polarization direction of the laser electric field. For a laser field linearly polarized in the $x$ direction, $\mathbf{e}_{\bot}=\mathbf{e}_x$,  while for a circularly polarized one, $\mathbf{e}_{\bot}=1/\sqrt{2}(\mathbf{e}_x\pm \mathrm{i} \mathbf{e}_y)$.

\begin{figure}
\begin{center}
\includegraphics[width=10cm]{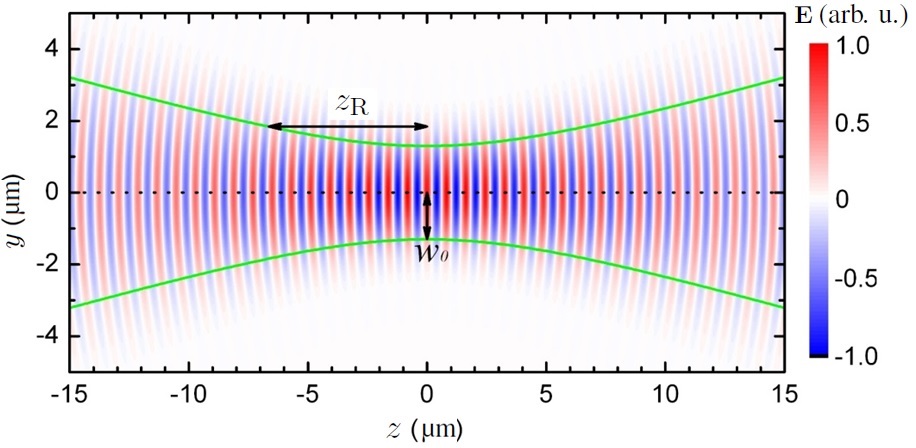}
\caption{Electric field of a Gaussian beam in the ($y,z$) plane; the focal plane is located at $z=0$}
\label{fig:gaussbeam}
\end{center}
\end{figure}

The propagation of the Gaussian laser pulse is fully characterized by the beam waist $w(z)$, the radius of curvature of the wavefront $R(z)$, and the Gouy phase shift $\psi_\mathrm{g}(z)$. As illustrated in Fig.\ \ref{fig:gaussbeam}, these parameters evolve along the $z$ axis as
\begin{equation}
\begin{split}
w(z)&=w_0\sqrt{1+\left(\frac{z}{z_\mathrm{R}}\right)^2}\ ,\\
R(z)&=z\left[1+\left(\frac{z}{z_\mathrm{R}}\right)^2\right]\ ,\\
\psi_g(z)&=\arctan\left(\frac{z}{z_\mathrm{R}}\right)\ ,
\end{split}
\label{eq:parameter}
\end{equation}
where $z_\mathrm{R}=\pi w^2_0/\lambda_0$ is the Rayleigh length, which represents the position at which the laser beam transverse area is doubled, compared with the one in the focal plane, owing to diffraction. The beam divergence far from the focal plane  ($z\gg z_\mathrm{R}$) is given approximately by $\theta_\mathrm{L}\simeq\lambda_0/\pi w_0$.

The electromagnetic fields $\mathbf{E}$ and $\mathbf{B}$ can be expressed in terms of the scalar potential $\Phi$ and the vector potential $\mathbf{A}$,
\begin{equation}
\begin{split}
\mathbf{E}&=-\nabla\Phi-\frac{\partial\mathbf{A}}{\partial t}\ ,\\
\mathbf{B}&=\nabla\times\mathbf{A}\ ,
\end{split}
\label{eq:potential}
\end{equation}
with the Coulomb gauge $\nabla\cdot\mathbf{A}=0$. In a vacuum $\Phi=0$, and both fields, $\mathbf{E}$ and $\mathbf{B}$,  depend only on the vector potential $\mathbf{A}$.
It is useful to define a  normalized peak vector potential, also called the laser strength, $a_0=eA_0/(mc)=eE_\mathrm{L}/(mc\omega_0)$.

The quantities usually measured during experiments are the laser energy,  $\mathcal{E}_\mathrm{L}$, and its distribution in space, and the pulse duration, $\tau_0$; they are used to evaluate the laser intensity, which is a key parameter for laser plasma interaction.
The laser  intensity is defined as
\begin{equation}
I_\mathrm{L}=c^2\varepsilon_0 \langle\mathbf{E}\times\mathbf{B}\rangle=\frac{c\varepsilon_0}{2}\left|\mathbf{E}\right|^2 \ ,
\label{eq:intensity}
\end{equation}
where $\varepsilon_0=8.85\times10^{-12}\UF/\UmZ$ is the permittivity of free space.
The laser power for a Gaussian pulse in time is written as
\begin{equation}
P=2\sqrt{\frac{\ln(2)}{\pi}}\frac{\mathcal{E}_\mathrm{L}}{\tau_0}\simeq\frac{\mathcal{E}_\mathrm{L}}{\tau_0}\ ,
\label{eq:power}
\end{equation}
and the corresponding peak laser intensity in the focal plane is
\begin{equation}
I_0=\frac{2P}{\pi w_0^2}\simeq\frac{2\mathcal{E}_\mathrm{L}}{\pi\tau_0w_0^2}\ .
\label{eq:pintensity}
\end{equation}
Equation (\ref{eq:pintensity}) shows that the peak intensity can be calculated from the measurements of energy, duration, and spot size, provided the shape of the laser beam is known.
Substituting Eq.\ (\ref{eq:intensity}) into the expression $a_0=eE_\mathrm{L}/(mc\omega_0)$, $a_0$ can be expressed as a function of the intensity in practical units,
\begin{equation}
\label{eq:a0I}
a_0=\sqrt{\frac{e^2}{2\pi^2\epsilon_0m^2_ec^5}\lambda_0^2I_0}\simeq0.86\lambda _0\ [\Uum]\ \sqrt{I_0\ \left[10^{18}\UW/\UcmZ^2\right]}\ .
\end{equation}

\begin{figure}
\begin{center}
\includegraphics[width=14.5cm]{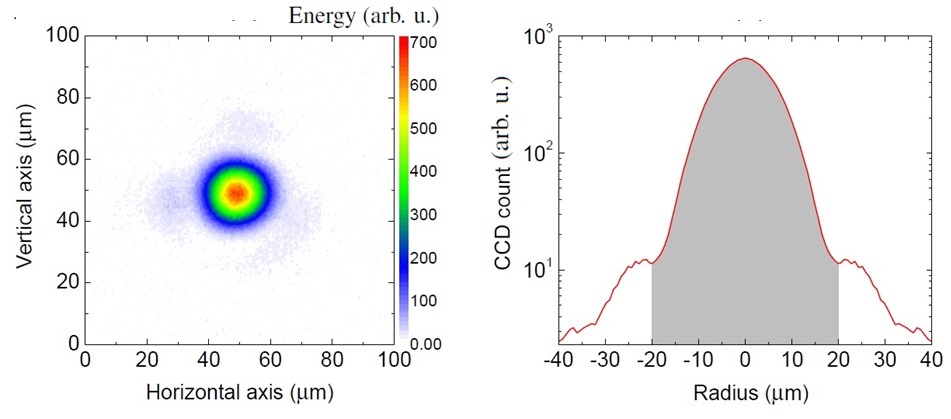}
\caption{ Left: energy distribution in the focal plane. Right: corresponding averaged radial profile of laser
energy in logarithmic scale; the grey shaded area, with a boundary at the first minimum of the focal spot, contains about 84\% of laser energy in the focal plane.}
\label{fig:focalspot}
\end{center}
\end{figure}

An example of energy distribution in the focal plane, measured at the Lund Laser Centre during an experiment \cite{jupop13}, is shown in Fig.\ \ref{fig:focalspot}. As is often the case, before focusing, the energy delivered by the laser system exhibits  a nearly flat-top cylindrically symmetrical distribution in the transverse plane. Therefore, in the focal plane, the corresponding energy distribution is not purely Gaussian, and it can be seen from the right of Fig.\ \ref{fig:focalspot}  that it exhibits a profile close to an Airy pattern. The radial profile is obtained by averaging the energy distribution from the left-hand image over the azimuthal angle. The focal spot shown in Fig.\ \ref{fig:focalspot} was achieved after optimization of the symmetry of the distribution,  by tuning a deformable mirror placed after the compressor, to compensate for aberrations in the laser wavefront. The average radius of the focal spot at first minimum can be determined from the radial profile;  in this example, it was measured as $(19.7 \pm 0.8 )\Uum$, which yields an on-axis peak intensity of $(5.4 \pm 0.1) \times 10^{18}\UW/\UcmZ^2$ and a normalized laser vector potential $a_0 = 1.6$. The energy fraction contained within the grey shaded area in Fig.\ \ref{fig:focalspot} is estimated to equal $84\%$ of the energy in the focal plane.

\subsection{Electron motion in a laser field}

For currently used laser systems, the maximum intensity usually exceeds $10^{18} \UW/\UcmZ^2$, corresponding to an electric field amplitude larger than $10^{12}\UV/\UmZ$. As the laser field is transverse to the direction of propagation,  single electrons mainly wiggle in this field, and it cannot be used directly to accelerate electrons in free space.

The motion of a single electron with charge $-e$ and mass $m_\mathrm{e}$ in the laser fields $\mathbf{E}$ and $\mathbf{B}$ is described by the Lorentz equation,
\begin{equation}
\frac{\mathrm{d}\mathbf{p}}{\mathrm{d}t}=-e(\mathbf{E}+\mathbf{v}\times\mathbf{B}) \, ,
\label{eq:lorentz}
\end{equation}
where $\mathbf{p}=\gamma m_\mathrm{e}\mathbf{v}$ is the electron momentum,  $\gamma=(1-\beta^2)^{-1/2}$  is the relativistic factor, and $\beta~=~v/c$ denotes the normalized velocity. Approximating the laser field as a plane electromagnetic wave polarized along the $x$ axis and propagating along the $z$ axis,  $\mathbf{E}(z)=E_\mathrm{L}\cos(k_0z-\omega_0t)\mathbf{e}_x$, or in terms of vector potential using Eq.\ (\ref{eq:potential}): $\mathbf{A}(z)=A_0\sin(k_0z-\omega_0t)\mathbf{e}_x$, with $A_0=E_\mathrm{L}/\omega_0$. Taking into account  $\left|\mathbf{B}\right|~=~\left|\mathbf{E}\right| /c$, the second term  in  the right-hand side of Eq.\ (\ref{eq:lorentz}) can be neglected in the non-relativistic regime, when $\beta\ll1$. Then Eq.\ (\ref{eq:lorentz}) becomes simply
\begin{equation}
\frac{\mathrm{d}\mathbf{p}}{\mathrm{d}t}=-e\mathbf{E}=e\frac{\partial \mathbf{A}}{\partial t}\ .
\label{eq:lorentz1}
\end{equation}
Therefore, an electron, initially at rest at $z=0$, oscillates in the direction of the electric field with a  velocity
\begin{equation}
\beta=-\frac{eA_0}{mc}\sin(\omega_0t)\triangleq-a_0\sin(\omega_0t) \, .
\label{eq:velocity}
\end{equation}
It follows from Eq.~(\ref{eq:velocity})  that an electron initially at rest will oscillate in the laser field with no net energy gain.

For $a_0\gtrsim1$, the electron oscillation velocity will approach $c$, and the $\mathbf{v}\times\mathbf{B}$ component in the Lorentz equation must be taken into account. The solution of  Eq.~(\ref{eq:lorentz}) in the relativistic regime can be found,  for example, in Ref. \cite{gibbon05}.
In the frame co-moving with the laser pulse, $\mathbf{a}(z)=a_0\sin(k_0\xi)\mathbf{e}_x$, where $\xi=z-ct$ is the coordinate in the frame co-moving with the laser. The normalized momentum of the electron can be written as
\begin{equation}
\begin{split}
u_x&=\gamma\beta_x=\frac{\mathrm{d}x}{\mathrm{d}\xi}=a=a_0\sin(k_0\xi)\ , \\
u_z&=\gamma\beta_z=\frac{\mathrm{d}z}{\mathrm{d}\xi}=\frac{a^2}{2}=\frac{a^2_0}{2}\sin^2(k_0\xi)\ .
\end{split}
\label{eq:momentum}
\end{equation}
The electron velocity is always positive in the $z$ direction, so that the $\mathbf{v}\times\mathbf{B}$ force pushes the electron  forward. Integrating Eq.\ (\ref{eq:momentum}) gives the electron coordinates along the trajectory:
\begin{equation}
\begin{split}
x&=-\frac{a_0}{k_0}\cos(k_0\xi)\ ,\\
z&=\frac{a^2_0}{8k_0}\left[2k_0\xi-\sin(2k_0\xi)\right]\ .
\end{split}
\label{eq:trajectory}
\end{equation}

\begin{figure}
\begin{center}
\includegraphics[width=10cm]{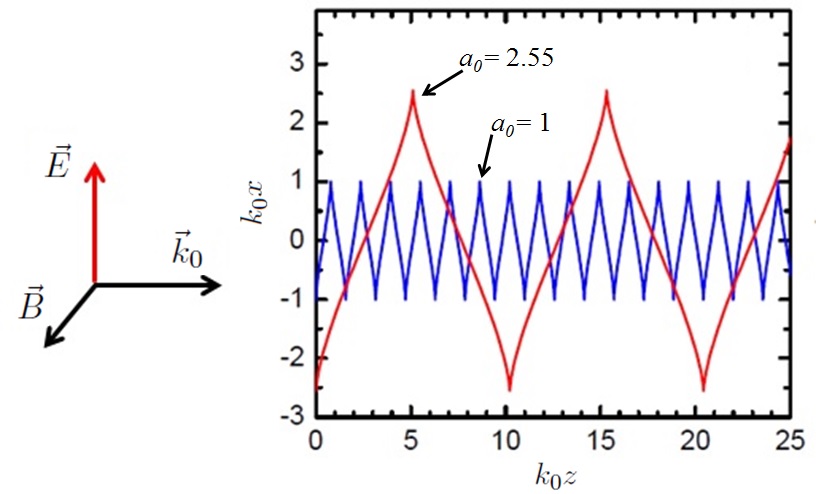}
\caption{Trajectory of an electron in the laser transverse field in the laboratory frame for two values of the laser strength.}
\label{fig:ewiggle}
\end{center}
\end{figure}

The set of Eq.\ (\ref{eq:trajectory}) indicates that the electron not only moves forward but also oscillates at  twice the laser frequency in the longitudinal $z$ direction. Figure \ref{fig:ewiggle} shows the electron trajectories for two values of $a_0$. The longitudinal momentum scales with the square of the laser strength as $a^2_0$, while the transverse one linearly depends on the laser strength by $a_0$. Hence, for $a_0\gg1$,  the longitudinal motion of the electron  dominates  the transverse oscillation.
The excursion distances along the
$x$ and $z$ axis calculated over one period become equal for $a_0=8/\pi\simeq2.55$.
For a laser wavelength of 800 nm, $a_0=2.55$ corresponds to $I_0=1.4\times10^{19}$ W/cm$^2$.

This simple analysis  shows the importance of operation in the relativistic regime, to achieve elec\-tron motion along the laser propagation direction. Nevertheless, as
shown in Fig.~\ref{fig:ewiggle}, the electron is merely pushed forward during each light cycle but does not gain energy from the laser. The overall net effect is only to transfer the electron to a new longitudinal position.
However,   in this section,  the laser beam was assumed to be a uniform and infinite plane wave. This is generally not the case in experiments, where  laser pulses need to be  tightly focused to achieve high intensity. The laser intensity is not uniform but Gaussian-like in the transverse plane.
As shown next, the ponderomotive force associated with
the laser intensity gradient  excites a plasma wave, as the plasma plays the role of a transformer, to transfer laser energy to electrons.

\subsection{Electromagnetic waves in plasmas}

In a plasma,  electrons oscillate around an equilibrium position with a characteristic frequency defined as the plasma frequency \cite{chen83}
\begin{equation}
\omega_\mathrm{p}=\sqrt{\frac{n_\mathrm{e}e^2}{m_\mathrm{e}\varepsilon_0}} \ .
\end{equation}
 In the non-linear regime, the plasma frequency is modified by  relativistic effects to $\omega_{\mathrm{pNL}}=\omega_\mathrm{p}/\sqrt{\gamma}$.
The dispersion relation of an electromagnetic wave in a  plasma can be written as
\begin{equation}
\label{eq:dispersion}
\omega_0^2=\omega_\mathrm{p}^2+c^2k^2\ .
\end{equation}
For $\omega_0>\omega_\mathrm{p}$, $k$ is real and the wave can propagate in the plasma; $k$ becomes imaginary for $\omega_0<\omega_\mathrm{p}$, and the wave is
evanescent. The light is thus either transmitted, damped, or reflected in the plasma, depending on the plasma density. The critical density, $n_\mathrm{c}$, is defined as the density for which $\omega_0=\omega_\mathrm{p}$, and can be written in practical units as
\begin{equation}
n_\mathrm{c}\ [10^{21}\Ucm^{-3}]=\frac{\omega^2_0m_\mathrm{e}\varepsilon_0}{e^2}=\frac{1.12}{\lambda^2_0\ [\Uum]}\  .
\end{equation}
The critical density corresponding to a laser wavelength of $0.8\Uum$ is $1.75\times10^{21}\Ucm^{-3}$. For $n_\mathrm{e}<n_\mathrm{c}$, the plasma is called underdense, and it is called overdense for $n_\mathrm{e}>n_\mathrm{c}$. The laser-driven wakefield relies on the excitation of a plasma wave in the underdense regime.

 The phase and group velocities of the electromagnetic field in the plasma are calculated from Eq.~\ (\ref{eq:dispersion}):
\begin{equation}
\label{eq:vph}
\begin{split}
v_{\mathrm{ph}}&=\frac{\omega_0}{k}=\sqrt{c^2+\frac{\omega_\mathrm{p}^2}{k^2}} \ , \\
v_\mathrm{g}&=\frac{\mathrm{d}\omega_0}{\mathrm{d}k}=\frac{c^2}{v_{\mathrm{ph}}}=\frac{c^2}{\sqrt{c^2+\omega_\mathrm{p}^2 /k^2}}\ .\\
\end{split}
\end{equation}

If the evolution of the driving laser pulse in the plasma is not significant during propagation, the phase velocity of the plasma wave is equal to the group velocity of the driving laser \cite{esarey96}. The normalized phase velocity and relativistic factor of the plasma wave are then given by
\begin{equation}
\begin{split}
\beta_\mathrm{p}&=\frac{v_\mathrm{g}}{c}=\sqrt{1-\frac{n_\mathrm{e}}{n_\mathrm{c}}} \, , \\
\gamma_\mathrm{p}&=\frac{1}{\sqrt{1-\beta_\mathrm{p}^2}}=\sqrt{\frac{n_\mathrm{c}}{n_\mathrm{e}}} \, .\\
\end{split}
\label{eq:betap}
\end{equation}
It can be seen that a lower density leads to  a higher  phase velocity and a larger relativistic factor.

\subsection{Plasma wave excitation}

In a plasma, the action of the ponderomotive force leads to the excitation of a plasma wave.
 This force is associated with the second-order electron motion [see Eq.\ (\ref{eq:lorentz})], averaged over a time-scale longer than the laser period.
 The three-dimensional (3D) ponderomotive force \cite{kruer03} for an electron can be written as
 \begin{equation}
 \mathbf{F}_\mathrm{p}=-m_\mathrm{e}c^2\nabla \left \langle  a^2 /2 \right \rangle =-m_\mathrm{e}c^2\nabla a_0^2 /2\ .
 \end{equation}
The ponderomotive force can be viewed as the radiation pressure of laser intensity. This force expels charged particles out of the region of high laser intensity, and  does not depend on the sign of the charged particle. Furthermore, it is inversely proportional to  particle mass $F_\mathrm{p}\propto 1/m$. Hence, under the same laser field, the acceleration exerted on a proton is only 10$^{-6}$ times that exerted on an electron, so that ion motion can be neglected for sufficiently low laser strength.

\begin{figure}
\begin{center}
\includegraphics[width=8cm]{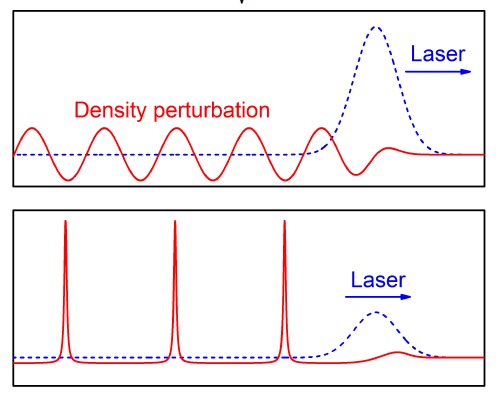}
\caption{Laser envelope (blue dashed line) and density perturbation (red solid line) along the axis of propagation in (top) the linear and (bottom) non-linear regimes.}
\label{fig:wakefield}
\end{center}
\end{figure}

Figure~\ref{fig:wakefield} illustrates the excitation of a plasma wave (red solid curve) in the linear (top graph) and non-linear (bottom graph) regimes, where the on-axis density exhibits localized spikes separated by a distance longer than the linear wavelength; the laser pulse envelope is indicated by a blue dashed line. The longitudinal electric field associated with a plasma wave is thus a space charge field linked to the periodic distribution of charges oscillating behind the laser pulse. The oscillation length is the plasma wavelength linked to the electron density of the plasma, $n_\mathrm{e}$, by the relation
\begin{equation}
 \lambda_\mathrm{p}\ [\UumZ] \simeq 33 \times (n_\mathrm{e}\ [10^{18}\Ucm^{-3}])^{-1/2} .
\end{equation}
 This plasma wave is called a relativistic plasma wave because its phase velocity, given by Eq.\ (\ref{eq:betap}), is  of the order of the laser group velocity  in the medium
when the plasma frequency $\omega _\mathrm{p}$ is much smaller than the laser frequency  $\omega _0$.
The amplitude of the longitudinal accelerating field can be written as:
\begin{equation}
E_\mathrm{p}\ [\UV[G]/\UmZ] \simeq 96 (n_\mathrm{e}\ [10^{18}\Ucm^{-3}])^{1/2} \frac {\delta n_\mathrm{e}}{n_\mathrm{e}}.
\end{equation}
The amplitude of this field is maximum for a density perturbation of 100\%, at the upper limit of the linear regime, where the plasma wave breaks.

\subsection{Regimes of the laser wakefield}

\begin{figure}
\begin{center}
\includegraphics[width=14cm]{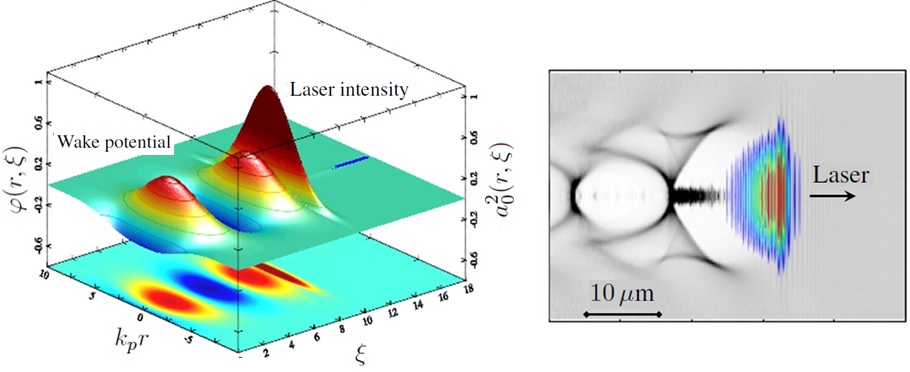}
\caption{ Illustration of the two regimes of the laser wakefield. Left: Excitation of a plasma wave in the linear regime; 3D view and projection in the horizontal plane of the normalized laser intensity, $ a_0 $, and wake potential, $ \varphi $. Right: Excitation of a plasma wave in the non-linear regime; map of density in the horizontal plane (grey scale, white is zero) and superposition of laser amplitude (colour scale, red is a maximum).}
\label{fig:2LPAregimes}
\end{center}
\end{figure}

Among the parameters determining the characteristics of accelerated electrons in a plasma are the laser characteristics (amplitude, transverse, and longitudinal sizes) and plasma properties, such as the electron density and distribution.
The laser strength $a_0$  is mainly used to distinguish between linear (or quasi-linear) and non-linear  regimes of laser wakefield excitation. Nevertheless, as the non-linear evolution of the laser includes a deformation of the spatial volume delimiting the most intense fraction of the laser pulse, the initial transverse size of the laser, $r_\mathrm{L}$, is also a key parameter.
In the quasi-linear regime, $a_0 \simeq 1 $ and
\begin{equation}
\frac{ k_\mathrm{p} ^2 r_\mathrm{L} ^2}{2}  > \frac{a_0 ^2}{\gamma _{\perp}}\ ,
\end{equation}
with $ \gamma _{\perp} = (1 + a_0 ^2 / 2)^{1/2}.$ The bubble or blow-out regime occurs for $a_0 > 1 $ and is characterized by
\begin{equation}
k_\mathrm{p} r_\mathrm{L} \leq 2 \sqrt{a_0}\ .
\end{equation}
Figure \ref{fig:2LPAregimes} illustrate the main features of  these two different regimes with simulation results. The left panel of Fig.\ \ref{fig:2LPAregimes} shows that the laser wakefield in the linear regime exhibits a regular oscillating behaviour \cite{andreev92}. The ponderomotive force varies as the laser energy gradient and creates a density distribution in the longi\-tudinal and transverse directions. The associated  transverse and longitudinal fields can be con\-trolled independently by adjusting the focal spot transverse size and the pulse duration. The accelerating structure is shaped as a sine wave with wavelength $ \lambda _\mathrm{p}$, typically in the range  10--100$\Uum$; its value is ad\-justed by tuning the plasma electron density.
The accelerating field is typically in the range 1--10$\UV[G]/\UmZ$, this value is limited to the wavebreaking field for a non-relativistic cold plasma, $E_0 = m_\mathrm{e} c \omega_\mathrm{p} /e$, and is of the order of $96\UV[G]/\UmZ$ for a plasma electron density $n_\mathrm{e} = 10^{18}\, $cm$^{-3}$. Wavebreaking is char\-acterized by the fact that electron oscillations become so large that the electrons can escape the collective motion. This may be at the origin of electron injection in the non-linear regime. In the linear regime,  wave\-breaking does not take place, and relativistic electrons need to be produced by an external source and injected into the linear plasma wave to be accelerated.

The right panel of Fig.\ \ref{fig:2LPAregimes} illustrates the main features of the non-linear regime. As the laser propa\-gates into the plasma, its front edge diffracts while the rest of the pulse self-focuses. Electrons are expelled by the ponderomotive force from the high-intensity volume and a plasma cavity, the white area in Fig.~\ref{fig:2LPAregimes}, which is void of electrons, is left behind the laser pulse. The expelled electrons travel along the field lines and accumulate at the back of the cavity, where they can be injected and accelerated. The cavity size is typically of the order of $10\Uum$, and produces accelerated electron bunches with a transverse size of the order of a few micrometres, owing to the focusing field inside the cavity. This regime is also called the blow-out or bubble regime \cite{pukhov02, lu2007}.

\subsection{Laser modulation in plasma}

The optical properties of the plasma are modified during the excitation of a plasma wave, and  can, in turn, modify the driving pulse. For the range of intensities of interest for a laser-driven  wakefield, there are two important effects, self-focusing and self-compression, corresponding to changes in the transverse size and duration of the laser pulse during its propagation.

\subsubsection{Self-focusing}

The propagation of a laser pulse in a plasma can be investigated in terms of  the spatial refractive index, $\eta (r)=c/v_{\mathrm{ph}}$. Recalling the definition of $v_{\mathrm{ph}}$ in Eq.\                                                     (\ref{eq:vph}), the refractive index for an underdense plasma, with uniform density $ n_{\mathrm{e}0}$,  and a large amplitude plasma wave, $ \omega_\mathrm{p} ^2 (r)= (\omega _{\mathrm{p}0}^2/\gamma) n_\mathrm{e}(r)/n_{\mathrm{e}0}$, is
\begin{equation}
\label{eq:fraction}
\eta(r)=\frac{c}{v_{\mathrm{ph}}}=\left(1-\frac{\omega_\mathrm{p}^2(r)}{\omega_0^2}\right)^{1/2}\simeq1-\frac{1}{2}\frac{n_\mathrm{e}(r)}
{\gamma(r)n_\mathrm{c}}\ .
\end{equation}
Equation (\ref{eq:fraction}) shows  that the spatial profile of the refractive index $\eta (r)$ can be modified by the relativ\-istic factor $\gamma(r)$ or the density distribution $n_\mathrm{e}(r)$. In the weakly relativistic case, Eq.\ (\ref{eq:fraction}) can be expanded as \cite{mori1997,esar09}
\begin{equation}
\label{eq:fraction2}
\eta(r) =1-\frac{\omega_\mathrm{p}^2}{2\omega_0^2}\left(1-\frac{a^2}{2}+\frac{\Delta n_e}{n_0} + \frac{\delta n_\mathrm{e}}{n_0}\right)\ ,
\end{equation}
where $\Delta n_\mathrm{e} / n_0$ takes into account the contribution of a preformed  plasma channel along the radius; the $\delta n_\mathrm{e}/n_0$  term is responsible for plasma wave guiding, self-channelling, and the self-modulation of long pulses;  the $a^2/2$ term corresponds to the contribution of relativistic laser guiding.

For a Gaussian pulse, with intensity peaked on-axis, $\partial a^2(r)/\partial r<0$, which satisfies the condition for refractive guiding,  $\partial \eta (r)/\partial r<0$ or $\partial v_{\mathrm{ph}}(r)/\partial r>0$. This implies that the on-axis phase velocity is less than the off-axis velocity, making the laser wavefront curved. The plasma thus plays the role of a convex lens, focusing the laser beam towards the propagation axis. This mechanism is known as self-focusing, and is able to balance laser  diffraction. As shown in Eq.\ (\ref{eq:fraction2}), the modulation of the refractive index leading to
beam focusing scales with $a^2$. Therefore, for a given beam divergence, there is a minimum threshold for laser intensity to balance diffraction. Self-focusing will occur when the laser
power $P$ exceeds a critical power $P_\mathrm{c}$, which can be written as \cite{sun1987,sprang1988,esarey1993}
\begin{equation}
\label{eq:Pc}
P_\mathrm{c}=\frac{8\pi\varepsilon_0m_\mathrm{e}^2c^5\omega_0^2}{e^2\omega_\mathrm{p}^2}\simeq17\frac{\omega_0^2}{\omega_\mathrm{p}^2} \ [\UGW]\ .
\end{equation}
For example, the critical power for a $\lambda=0.8\Uum$ laser  at plasma density $n_0=7\times10^{18}\Ucm^{-3}$ is $P_\mathrm{c}=4.25\UW[T]$.

The radial ponderomotive force expels electrons from the axis, thus creating a radial density gradi\-ent $\partial(\delta n_\mathrm{e}(r))/\partial r>0$, and a negative transverse gradient of the refractive index $\partial \eta (r)/\partial r<0$, act\-ing to focus the laser beam, although for $P\ll P_\mathrm{c}$,  the ponderomotive force is not sufficient to guide the laser \cite{esarey96}. When the laser power approaches the critical power, relativistic self-guiding dominates  laser propagation, while ponderomotive channelling  enhances self-guiding. The contribution of pondero\-motive channelling slightly relaxes the critical power for self-guiding to $P_\mathrm{c}=16.8(\omega_0^2/\omega_p^2)\ [\UW[G]]$.

\subsubsection{Self-compression}
In addition to the beam size evolution in the transverse direction,  the variation of plasma density along the propagation axis modulates the laser pulse longitudinally. The density variation along the propagation axis $\delta n_\mathrm{e}(\xi)$ during the laser pulse makes it experience different local refraction indices, and consequently compress or stretch in the temporal domain.

Using Eq.~(\ref{eq:fraction2}), the local group velocity of the drive laser, or the phase velocity of the plasma wave, can be examined, taking into account the longitudinal dependence of $\eta$:
\begin{equation}
\label{eq:compress}
v_\mathrm{g}/ c \simeq \eta \simeq 1-\frac{\omega_\mathrm{p}^2}{2\omega_0^2}\left(1-\frac{a^2}{2}+\frac{\delta n_\mathrm{e}}{n_0}\right)\ .
\end{equation}
$\delta n_\mathrm{e}=0$ gives the laser group velocity in the background plasma: $v_{\mathrm{g}0}$. For positive density variation $\delta n_\mathrm{e}>0$, the laser group velocity decreases $v_\mathrm{g}-v_{\mathrm{g}0}<0$, whereas  $\delta n_\mathrm{e}<0$ corresponds to an increase in laser group velocity: $v_\mathrm{g}-v_{\mathrm{g}0}>0$. As a result of plasma wave excitation (see Fig.\ \ref{fig:wakefield}), the  rear of the laser pulse moves faster than its front edge. The laser pulse is temporally compressed; this is also called pulse shortening. This effect plays an important role in laser-driven wakefields as it contributes to matching the laser pulse duration and plasma wave period.

As a  consequence of temporal compression, the spectral bandwidth of the short laser pulse will increase.
Most of the laser pulse can be redshifted, while strong  blueshifts occur at the back of the pulse for pulses longer than the plasma period \cite{schr2010}. As the  spectral modulation of the driving laser is created by the plasma modulation, this  provides a good means of diagnosing the excited plasma wave \cite{wojda2009}.

\subsection{Energy gain of an electron in a plasma wave}

The energy gain, $ \Delta W$, of an electron accelerated in a plasma wave is proportional to the product of the accelerating longitudinal field associated with the plasma wave, $ E_\mathrm{p}$, and the length, $L_{\mathrm{acc}}$, over which the electron is submitted to this field,
\begin{equation}
\Delta W = e E_\mathrm{p} L_{\mathrm{acc}}\ .
\end{equation}
The amplitude of the accelerating field and the length of acceleration depend on the regime of the laser wakefield and can be optimized in different ways.
Three main mechanisms may limit the acceler\-ation distance: laser diffraction  typically  limits acceleration to  the Rayleigh length; pump depletion defines the length over which half of the laser energy is transferred to the plasma wave; and de\-phasing, in which acceler\-ated electrons outrun the plasma wave and enter a decelerating phase, defines the de\-phasing length.
Diffraction is a limitation due to laser propagation in the medium and can be overcome in the non-linear regime by self-focusing: in that case, the non-linearities imposed on the medium can shape the transverse density profile to act as a transient focusing length. In the linear regime, where these non-linear effects are negligible, external guiding has to be implemented; the use of capillary tubes for this purpose will be described in Section \ref{sec:capillary tubes}. Electron dephasing and laser depletion are intrinsic to laser plasma acceleration and depend  on the electron density. Their dependences are described next.

\subsubsection{Dephasing}
As the phase of the plasma wave evolves at the group velocity of the laser in the plasma, a relativistic electron accelerated in this wave can explore different phases of the accelerating field. The dephasing length \cite{esarey96, geddes05a}, $L_\mathrm{d}$, is defined as the length over which an electron can be accelerated before reaching a decelerating period of the electric field.

\begin{figure}
\centering
\includegraphics[width=8cm]{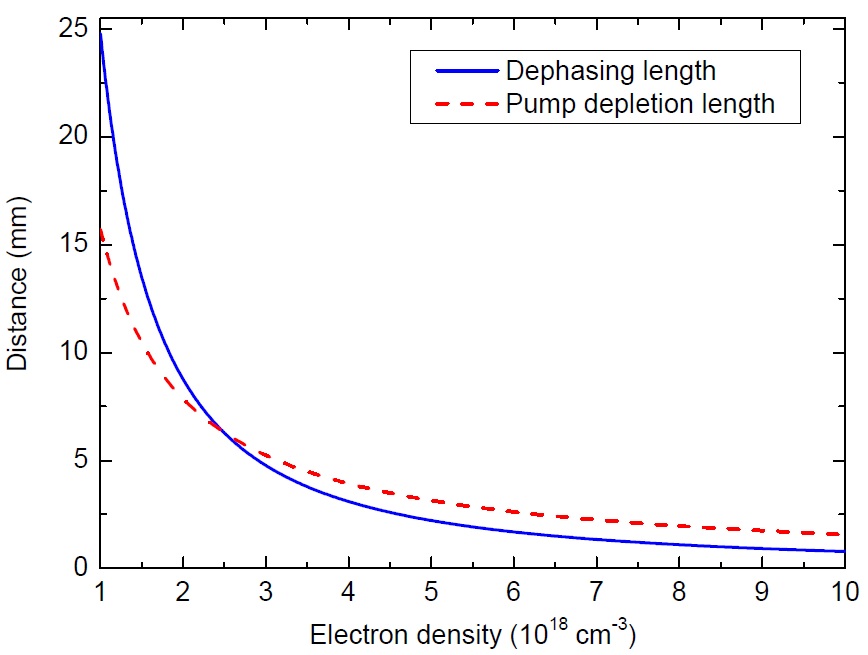}
\caption[3D dephasing length and pump depletion length]{3D dephasing length and pump depletion length as functions of  plasma electron density. The laser pulse has a duration of $30\Ufs$ and a peak intensity corresponding to $a_0=4$.}
\label{fig:Lpd}
\end{figure}

In the one-dimensional (1D) linear regime, $L_\mathrm{d}$ is evaluated as follows: in the co-moving frame, the maximum length over which the field is accelerating during one period is $\lambda_\mathrm{p}/2$. Assuming that the accelerated electron propagates at approximately $c$, and keeping in mind that the plasma wave phase moves forward with velocity  $v_\mathrm{g}$, $L_\mathrm{d}^{\mathrm{1D}}$ is estimated as
\begin{equation}
L_\mathrm{d}^{\mathrm{1D}}=\frac{\lambda_\mathrm{p}}{2(c-v_\mathrm{g})}c=\frac{\lambda_\mathrm{p}}{2(1-\beta_\mathrm{p})}\simeq\frac{n_\mathrm{c}}{n_\mathrm{e}}\lambda_\mathrm{p}=\frac{\omega_0^2}{\omega_\mathrm{p}^2}\lambda_\mathrm{p}\ .
\end{equation}
This expression can be generalized to two-dimensional (2D), by noting that for a 2D plasma wave, there is only a quarter of the period ($\lambda_\mathrm{p}$/4) in which the longitudinal electric field is accelerating and the radial field is focusing. Therefore, the dephasing length becomes
\begin{equation}
L_\mathrm{d}^{\mathrm{2D}}=L_\mathrm{d}^{1D}/2=\frac{\omega_0^2}{2\omega_\mathrm{p}^2}\lambda_\mathrm{p}\ .
\end{equation}
In the 3D bubble regime, the distance in the co-moving frame for dephasing
becomes the bubble radius $R_\mathrm{b}$. The phase velocity of the plasma wave is modified to
 $\beta_\mathrm{p}=1-3\omega_\mathrm{p}^2/(2\omega_0^2)$. Accordingly, the dephasing length $L_\mathrm{d}^{\mathrm{3D}}$ is given by
\begin{equation}
L_\mathrm{d}^{\mathrm{3D}}=\frac{R_\mathrm{b}}{1-\beta_\mathrm{p}}\simeq\frac{2}{3}\frac{\omega_0^2}{\omega_\mathrm{p}^2}R_\mathrm{b}=\frac{4}{3}\frac{\omega_0^2}{\omega_\mathrm{p}^2}\frac{\sqrt{a_0}}{k_\mathrm{p}}\ .
\end{equation}
This expression shows that the 3D non-linear dephasing length depends on both the plasma electron density and the laser intensity. A longer $L_\mathrm{d}^{\mathrm{3D}}$ can be achieved either
decreasing the plasma density, as illustrated in Fig.\ \ref{fig:Lpd}, or by increasing the laser intensity.

\subsubsection{Pump depletion}

Another underlying limit is the so-called  pump depletion, associated with the length $L_{\mathrm{pd}}$, over which the driving laser becomes depleted.  As the laser pulse travels in the plasma, it transfers its energy to the plasma wave. The pump depletion length  is defined as the distance over which  the energy contained in the plasma
wave equals that of the driving laser, $ E_{\mathrm{p}} ^2 L_{\mathrm{pd}} = \int  E_\mathrm{L}^2 \mathrm{d}\xi$.
For a Gaussian driving laser pulse, the linear pump depletion length is
given by  \cite{shadwick09}
\begin{equation}
L_{\mathrm{pd}}^{\mathrm{L}}=\frac{\omega_0^2}{\omega_\mathrm{p}^2}\frac{c\tau_0}{a_0^2}\ .
\end{equation}
In the non-linear regime, the pump depletion length is estimated via the etching velocity $v_{\mathrm{etch}}\simeq c\omega_\mathrm{p}^2/\omega_0^2$ \cite{decker1996}; this describes the erosion velocity of the laser front that excites the plasma wave, before the start of self-focusing. The laser will be depleted over the depletion length given by
\begin{equation}
L_{\mathrm{pd}}^{\mathrm{NL}}=\frac{c}{v_{\mathrm{etch}}}c\tau_0=\frac{\omega_0^2}{\omega_\mathrm{p}^2}c\tau_0\ .
\end{equation}
Lu \textit{et al.}\cite{lu2007} showed  that this expression of non-linear pump depletion length $L_{\mathrm{pd}}^{\mathrm{NL}}$ is valid for 2D and 3D non-linear cases.

Both the 3D dephasing length $L_\mathrm{d}^{\mathrm{3D}}$ and the pump depletion length $L_{\mathrm{pd}}^{\mathrm{NL}}$  are plotted as  functions of plasma density, for  $a_0=4,$ in Fig.~\ref{fig:Lpd}.   Typically, the two lengths are of the order of a few millimetres in the density range above  $n_\mathrm{e}=3\times10^{18}\Ucm^{-3}$, where dephasing  dominates pump depletion. The most efficient use of laser energy is achieved around the density where $L_\mathrm{d}^{\mathrm{3D}}\simeq L_{\mathrm{pd}}^{\mathrm{NL}}$.  Below $n_\mathrm{e}=2.5\times10^{18}\Ucm^{-3}$,  pump depletion occurs over a shorter length than dephasing, and both lengths increase quickly when the density becomes of the order of  $1 \times10^{18}\Ucm^{-3}$ .

\subsubsection{Scaling laws}
\begin{figure}
\begin{center}
\includegraphics[width=10cm]{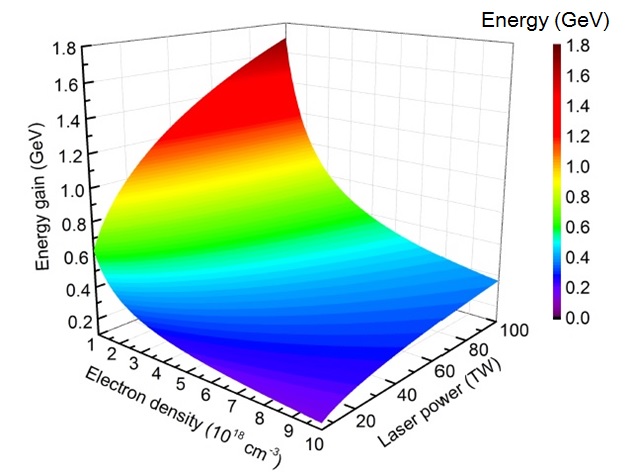}
\caption{Energy gain over the dephasing length as a function of plasma density and laser power, calculated using Eq.~(\ref{eq:Elu}).}
\label{fig:energygainnP}
\end{center}
\end{figure}

Scaling laws in the different regimes have been established from phenomenological considerations  by Lu \textit{et al.} \cite{lu2007}, and are summarized in Table \ref{tab:Lu}. These scaling laws play an important role in understanding laser plasma acceleration mechanisms, and have been verified in numerous experiments. Scaling laws can be used, for example, to predict the energy gain in the non-linear regime. Using an average value of the accelerating field in the bubble regime, and an average value for acceleration over the dephasing length, the energy gain can be approximately written as
\begin{equation}
\Delta\mathcal{E}\simeq\frac{2}{3} a_0 \frac{\omega_0^2}{\omega_\mathrm{p}^2}m_\mathrm{e}c^2\simeq m_\mathrm{e}c^2\left(\frac{e^2P}{m_\mathrm{e}^2c^5}\right)^{1/3}\left(\frac{n_\mathrm{c}}{n_\mathrm{e}}\right)^{2/3}\ ,
\end{equation}
which becomes, in  practical units,
\begin{equation}
\label{eq:Elu}
\Delta\mathcal{E}\ [\UGeV] \simeq1.7\left(\frac{P\ [\UWZ[T]]}{100}\right)^{1/3}\left(\frac{0.8}{\lambda_0\ [\UumZ]}\right)^{4/3}\left(\frac{1}{n_\mathrm{e}\ [10^{18}\Ucm^{-3}]}\right)^{2/3}\ .
\end{equation}
This expression shows that the plasma electron density is a key parameter for tuning the electron bunch energy: a large energy gain over the dephasing length in the 3D non-linear regime can be achieved for low electron density and large laser power.
Figure \ref{fig:energygainnP} shows the electron energy gain as a function of plasma density and laser
power, calculated from Eq.~(\ref{eq:Elu}). For a laser  power of 100\UW[T], Eq.~(\ref{eq:Elu}) predicts
that electrons can be accelerated up to $1.7\UGeV$ at a plasma density of  $n_\mathrm{e} = 1 \times 10^{18}\Ucm^{-3}$. The energy gain displays a stronger dependence on plasma density than on laser power. Hence, to achieve higher electron energy, a lower electron plasma density is desirable. Nevertheless, one must always remember the laser size matching condition, which requires $w_0 = \lambda_\mathrm{p} \sqrt{a_0}/ \pi$, which is proportional to $1/\sqrt{n_\mathrm{e}}$. When $n_\mathrm{e}$ is decreased, the laser power
must be increased accordingly to meet the requirement for producing a larger bubble, as $P = I_0 \pi w_0 ^2 /2$, which is proportional to $1/ n_\mathrm{e}$ .

\begin{table}
\centering
\caption[Summary of scaling laws for different regimes in \cite{lu2007}]{Summary of scaling laws for linear and non-linear regimes  \cite{lu2007}, for the dephasing length $L_\mathrm{d}$, the depletion length, $L_{\mathrm{pd}}$, the relativistic factor of the plasma wave $\gamma _\mathrm{p}$ , and $\Delta\mathcal{E}$, the energy gain of an electron accelerated over the dephasing length.}
\label{tab:Lu}
\renewcommand{\arraystretch}{2}
\begin{tabular}{ l  c  c  c  c  c  c }
\hline \hline
   & $a_0$ & $w_0$ & $L_\mathrm{d}$ & $L_{\mathrm{pd}}$ & $\gamma_\mathrm{p}$ & $\Delta\mathcal{E}/m_\mathrm{e}c^2$ \\
\hline
Linear & <1 & $\lambda_\mathrm{p}$ & $\frac{\omega_0^2}{\omega_\mathrm{p}^2}\lambda_\mathrm{p}$ & $\frac{\omega_0^2}{\omega_\mathrm{p}^2}\frac{c\tau_0}{a_0^2}$ & $\frac{\omega_0}{\omega_\mathrm{p}}$ & $a_0^2\frac{\omega_0^2}{\omega_\mathrm{p}^2}$ \\

1D non-linear & >1 & $\lambda_\mathrm{p}$ & $4a_0^2\frac{\omega_0^2}{\omega_\mathrm{p}^2}\lambda_\mathrm{p}$ & $\frac{1}{3}\frac{\omega_0^2}{\omega_\mathrm{p}^2}c\tau_0$ & $\sqrt{a_0}\frac{\omega_0}{\omega_\mathrm{p}}$ & $4a_0^2\frac{\omega_0^2}{\omega_\mathrm{p}^2}$  \\

3D non-linear & >2 & $\frac{\sqrt{a_0}}{\pi}\lambda_\mathrm{p}$ & $\frac{4}{3}\frac{\omega_0^2}{\omega_\mathrm{p}^2}\frac{\sqrt{a_0}}{k_\mathrm{p}}$ & $\frac{\omega_0^2}{\omega_\mathrm{p}^2}c\tau_0$ & $\frac{1}{\sqrt{3}}\frac{\omega_0}{\omega_\mathrm{p}}$ & $\frac{2}{3}a_0\frac{\omega_0^2}{\omega_\mathrm{p}^2}$  \\
\hline \hline
\end{tabular}
\end{table}

\begin{figure}
\begin{center}
\includegraphics[width=10.5cm]{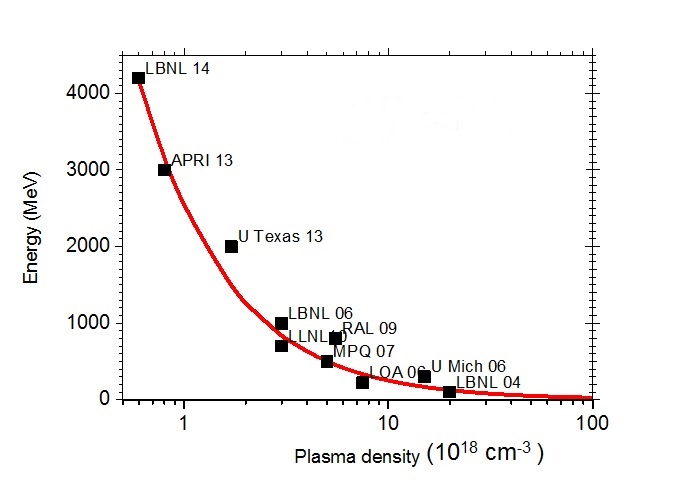}
\caption{Energy gain measured in various experiments around the world, plotted with black squares and labels, follow the dependence $\Delta W \propto 1/n$. Labels indicate institute acronyms and year of publication.}
\label{fig:summexpres}
\end{center}
\end{figure}

Numerous experimental results in the non-linear regime, where electrons from the plasma are self-injected into the accelerating structure, follow the scaling law
for energy gain $\propto 1/n_\mathrm{e}$, as illustrated in Fig.~\ref{fig:summexpres}.
Figure \ref{fig:summexpres} shows that when the density is divided by a factor of 10, from $ n_\mathrm{e} = 8\times 10^{18} \, $cm$^{-3}$ to $ n_\mathrm{e} = 0.8 \times 10^{18} \, $cm$^{-3}$, the maximum energy of the electron bunch is multiplied by the same factor. It should be noted that parameters other than the plasma density can change significantly between the different cases plotted in this graph: in particular, the laser power has been increased over the years and has contributed to the experimental observation of electrons in the gigaelectronvolt range for petawatt-class laser systems. The length of the plasma was also increased: starting from millimetre-scale gas jets at higher densities, gas cells of length between one and a few centimetres are used in  intermediate regimes; the largest energy gain to date was obtained using external guiding by a plasma channel \cite{leemans14}.

In conclusion, experimental results are well understood regarding the maximum electron energy dependence with electron plasma density and clearly indicate the way to increase the energy further. Reducing the density significantly increases the dephasing length: the next challenge will be to achieve metre-scale plasma lengths while preserving the laser intensity over the whole distance. Next, we exam\-ine some aspects of low-density operation of  laser wakefield excitation over large distances.

\section{External guiding with capillary tubes}\label{sec:capillary tubes}

\subsection{Choosing a guiding mechanism}

Several methods have been developed to guide the focused laser over a distance longer than the Rayleigh length. Among them, self-focusing  is one of the most commonly employed, owing to the simplicity of its implementation.
Relying on this scheme, a powerful laser, $P_\mathrm{L}>P_\mathrm{c}$, can be guided over the pump depletion length, typically several millimetres long (see Fig.\ \ref{fig:Lpd}). To achieve self-focusing,  the laser power must be in excess of the critical power, given by $P_\mathrm{c}\UW[G]=17\omega_0^2/\omega_\mathrm{p}^2$,
which increases rapidly when the plasma density is decreased: for a plasma density $n_\mathrm{e}=1\times10^{18}\Ucm^{-3}$, $P_\mathrm{c}$  equals 30\UW[T].

 \begin{figure}
\begin{center}
\includegraphics[width=10cm]{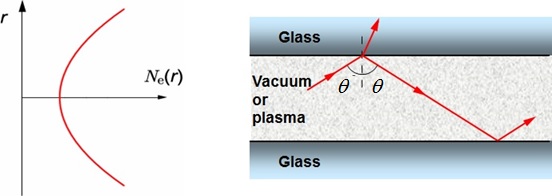}
\caption{Guiding methods: left, plasma channel, characterized by varying plasma density along the radius; right, capillary tube with glass wall, and laser reflection from the inner walls.}
\label{fig:extguiding}
\end{center}
\end{figure}

Self-focusing is the result of  a modification of the plasma density by the front of the laser pulse, acting to create a density structure able to compensate diffraction for the rear of the pulse.
A dens\-ity structure acting to compensate diffraction (see Eq.\ (\ref{eq:fraction2})) can be formed prior to the intense laser inter\-action.  A plasma channel  can be
created with the help of an external electrical discharge \cite{spence2000,spence2003} or by a heating laser pulse \cite{geddes05,leemans06}:  the hot on-axis plasma electrons move outward,
owing to radial hydrodynamic expansion, resulting in a density depletion on-axis and a nearly parabolic transverse dens\-ity profile, as illustrated in the left panel of Fig.~\ref{fig:extguiding}.     A parabolic plasma channel can guide a Gaussian beam with a constant spot size $w_0 = w_\mathrm{m}$, where $w_\mathrm{m}$ depends on the curvature of the channel. Under the matching condition,  lasers with intensity of the order of $10^{18}\UW/\UcmZ^2$  were successfully guided
by plasma waveguides over many Rayleigh lengths \cite{geddes05,butler2002}.
Good-quality guiding in plasma channels relies on fine tuning of
the incident laser parameters (spot
size, duration, energy)  with channel parameters. As
 the guiding mechanism relies on plasma density and its spatial profile,
this mechanism cannot be used for  applications requiring plasma parameters that are very different from those required for
laser guiding in the plasma channel.

When a capillary tube is used, as illustrated in the right panel of Fig.~\ref{fig:extguiding},  the laser beam is guided by  reflections at the inner capillary wall \cite{cros2002}, and laser guiding can  thus be achieved in vacuum or at low plasma density.  This guiding scheme does
not rely on laser power, or plasma density, and provides the opportunity to explore a large domain of laser plasma parameters.
 Laser guiding can, in principle, be achieved inside capillary tubes with total or partial reflection at the inner wall, depending on the material of the tube wall, which may be either a conductor or a dielectric material. Metallic tubes could be used to guide the laser beam without loss at the inner wall. In practice, their surface is usually not optically smooth for the laser wavelength and tube diameters used for laser guiding at high intensities. Dielectric capillaries, such as glass capillaries, are optically smooth and can be manufactured with a good reliability for a large range of inner diameters, wall thicknesses and lengths.

\subsection{Eigenmodes of capillary tubes}

\begin{figure}
\begin{center}
\includegraphics[width=12cm]{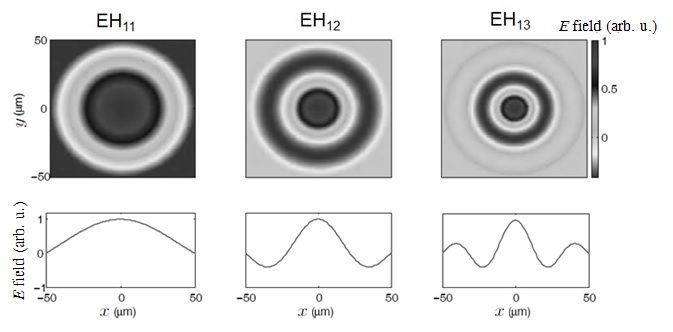}
\caption{Field distribution in the transverse plane for the three first hybrid modes EH$_{1m}$, plotted for $R_{\mathrm{cap}} = 50\Uum$;  bottom graphs show the corresponding distributions through $y=0$.}
\label{fig:modes123}
\end{center}
\end{figure}

Solving Maxwell's equations, with boundary conditions for dielectric surfaces at the capillary tube inner wall, gives hybrid mode solutions, with quasi-transverse electromagnetic modes.
 Solving the wave equa\-tion in cylindrical geometry, with boundary conditions describing the continuity of the field components at the boundary between the vacuum inside the tube and the dielectric wall, different families of  eigen\-modes
of capillary tubes are obtained \cite{cros2002}. An incident linearly polarized Gaussian laser beam can be efficiently coupled to the linearly polarized family of hybrid modes, namely the EH$_{1m}$ modes. The
transverse electric components of the EH$_{1m}$ modes at zero order can be found in Ref.~\cite{cros2002}.
For the EH$_{1m}$ modes, the transverse electric field amplitude inside the capillary tube can be written as
\begin{equation}
E_{1m}(r,z,t) = J_{0} (k_{\perp m}r)\exp(-k_m ^l z) \cos(\omega _0 t - k_{zm}z) \ ,
\label{eq:c4eigen}
\end{equation}
where $ k_{\perp m}$ is the transverse wavenumber of the mode with order $m$, defined as  $k_{\perp m} = (k_0 ^2 -  k_{zm} ^2)^{1/2} $, and is given by $k_{\perp m} = u_m / R_{\mathrm{cap}}$, $ R_{\mathrm{cap}}$ is the capillary tube inner radius,
$u_m$ is the $m\mathrm{th}$ root of $J_0(x)=0$, and $J_0$ is the Bessel function of integer order. Table \ref{tab:c4roots} gives the first nine values of $u_m$. $k_{zm}$ is the longitudinal wavenumber inside the tube, and $k_0$ is the wavenumber of the laser beam in free space.  Figure \ref{fig:modes123} illustrates the electric field amplitude in the transverse plane for the three first modes. The electric field of the  EH$_{1m}$ modes is maximum at the tube axis; for $m=1$, the field variation is close to the field of a Gaussian beam.

The exponential decay term in Eq.~(\ref{eq:c4eigen}) indicates that the electric field is damped along the dir\-ection of propagation, $z$: each reflection  at the dielectric capillary wall is associated with a refracted fraction of the laser beam inside the dielectric wall. This refracted fraction is minimum for the smallest perp\-endicu\-lar wavenumber, corresponding to the grazing incidence.
 The characteristic damping co\-efficient $k_m^l$ is given by \cite{cros2002}:
\begin{equation}
k_m^l=\frac{u_m^2}{2k_{z0}^2R^3_{\mathrm{cap}}}\frac{1+\varepsilon_\mathrm{r}}{\sqrt{\varepsilon_\mathrm{r}-1}}\ ,
\end{equation}
where $\varepsilon_\mathrm{r}$ is the dielectric constant of the wall.
This shows that $k^l_m$ strongly depends on the capillary radius, $R_{\mathrm{cap}}$, the wavelength of the incident laser beam, and the mode order, through $u_m$. Laser damping is usually described by the attenuation length $L^l_m$, defined as
\begin{equation}
L_m^l=\frac{1}{k_m^l}=\frac{2k_{z0}^2R^3_{\mathrm{cap}}}{u_m^2}\frac{\sqrt{\varepsilon_\mathrm{r}-1}}{1+\varepsilon_\mathrm{r}}.
\label{eq:c4Lm}
\end{equation}
After a propagation distance of $L^l_m$, the field magnitude decreases by a factor $1/e$ and the beam energy by a factor $1/e^2$, owing to refraction losses. For example, the values of damping length $L^l_m$ for  a Ti:sapphire laser  ($\lambda_0=800\Unm$) guided inside a  $ 50\Uum$ radius capillary tube are given in Table~\ref{tab:c4roots} for the nine first modes.  $L^l_m$ drops rapidly with increasing mode order, which means that higher-order modes are damped over shorter distances. Therefore, the use of the fundamental mode is preferable, to achieve  laser guiding over a long propagation distance.

\begin{table}
\centering
\caption[$u_m$, $L^l_m$, and $v_{gm}$ for the first nine eigenmodes]{Values of $u_m$, $L^l_m$, and $v_{\mathrm{g}m}$ for the first nine modes of a capillary: radius, $50\Uum$; laser wavelength, $\lambda_0=800\Unm$.}
\label{tab:c4roots}
\begin{tabular}{l d s d d }
\hline\hline
\multicolumn {1}{l}{$m$} & \multicolumn {1}{c} {$u_m$} & \multicolumn {1} {c} {$L^l_m (\UcmZ)$} & \multicolumn {1} {c} {$v_{\mathrm{g}m}/c$} & \multicolumn {1} {c} {$\mathcal{F}_m^{\mathrm{max}}$ (10$^{-5}$)}\\
\hline
1 & 2.404826 & 91.7 & 0.99998 & 2.034 \\
2 & 5.520078 & 17.4 & 0.9999 & 4.604 \\
3 & 8.653728 & 7.1 & 0.9998 & 7.2011 \\
4 & 11.79153 & 3.8 & 0.9995 & 9.8049 \\
5 & 14.93092 & 2.4 & 0.9993 & 12.4113 \\
6 & 18.07106 & 1.6 & 0.9989 & 15.0189 \\
7 & 21.21164 & 1.2 & 0.9985 & 17.6272 \\
8 & 24.35247 & 0.9 & 0.9981 & 20.2359 \\
9 & 27.49348 & 0.7 & 0.9975 & 22.8449 \\
\hline\hline
\end{tabular}
\end{table}

The  group velocity is determined from the dispersion relation of an electromagnetic wave in a capillary tube, $k_0^2=(k_z^2+k^2_{\bot m})$, for $k^2_{\bot m}\ll k_0^2$, as
\begin{equation}
v_{\mathrm{g}m}\simeq c\left(1-\frac{k^2_{\bot m}}{k_0^2}\right)^{1/2}\ .
\end{equation}
The fourth column in Table \ref{tab:c4roots} shows the values of $v_{\mathrm{g}m}$. It can be seen that the group velocity is close to the velocity of light in free space, and decreases as the mode order increases. This again supports the use a lower-order mode with higher group velocity,  corresponding to a higher phase velocity for the wakefield.

Another important  issue associated with the use of a capillary tube is the threshold of material damage at the inner wall where reflection occurs, which determines the ability of capillary tubes to guide intense lasers. To examine this, we define the normalized flux at the  capillary inner wall, $\mathcal{F}_m$, as the ratio of the radial component of the Poynting vector at $r=R_{\mathrm{cap}}$ to the longitudinal component of the on-axis Poynting vector; it is given by \cite{cros2002}
\begin{equation}
\mathcal{F}_m=\frac{k^2_{\bot m}}{k^2_0}J_1^2(k_{\bot m}R_{\mathrm{cap}})\frac{\cos^2\theta+\varepsilon_r\sin^2\theta}{\sqrt{\varepsilon_r-1}}\ .
\end{equation}
$\mathcal{F}_m$ depends on the azimuthal angle $\theta$ and mode order $m$. $\mathcal{F}_m$ is minimum for $\theta=0,\ \pi$ and maximum for $\theta=\pi/2,\ 3\pi/2$. It also depends on the mode eigenvalue and capillary radius through $k^2_{\bot m}$. Its  maximum value  should be below the threshold of material breakdown, to ensure laser guiding without non-linear interaction at the wall. As shown in Table \ref{tab:c4roots}, $\mathcal{F}_m^{\mathrm{max}}$ increases by one order of magnitude from the fundamental mode to the ninth mode. Once again, it emphasizes the advantage of  using the fundamental mode. For a glass capillary, the ionization threshold is of the order of $10^{14}\UW/\UcmZ^2$ for an $800\Unm$ laser pulse with duration shorter than or of the order of $100\Ufs$ \cite{du1994}. The maximum intensity on-axis guided by a capillary of $50\Uum$ radius on the fundamental EH$_{11}$ mode without wall ionization is thus of the order of $10^{19}\UW/\UcmZ^2$.

\subsection{ Mode coupling }

When a laser beam is focused at the entrance of a capillary tube,  its energy has to be coupled to the capillary eigenmodes  before propagation. In this section, we describe  the conditions of coupling for two kinds of laser energy distribution in the focal plane, located at the entrance of the capillary tube: a Gaussian beam and an Airy beam, both of which can be good approximations to describe the beams used in experiments. As shown in the previous section, monomode guiding with the fundamental EH$_{11}$ mode has interesting properties for laser wakefield excitation, so the condition for monomode coupling will be particularly discussed.

\subsubsection{Coupling of a Gaussian beam}

\begin{figure}
\begin{center}
\includegraphics[width=13cm]{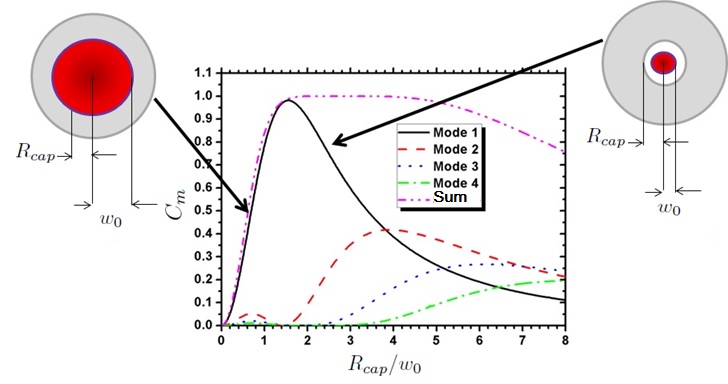}
\caption{Coupling coefficient of modes 1  to 4 for a Gaussian beam with waist $w_0$ incident at the entrance of a capillary with radius $R_{\mathrm{cap}}$ as a function of the ratio $R_{\mathrm{cap}} / w_0$.}
\label{fig:modecoupling}
\end{center}
\end{figure}

Laser beams can be described as transverse electromagnetic modes with a transverse Gaussian envelope. Assuming a linearly polarized Gaussian beam  focused at the capillary entrance ($z=0$), the amplitude of its electric field is written as $E_\mathrm{G}(r)=E_\mathrm{L}\exp(-r^2/w^2_0)$, while the electric field inside a capillary tube  is a superposition of the EH$_{1m}$ eigenmodes given by Eq.~(\ref{eq:c4eigen}). At the entrance ($z = 0$), the continuity of fields reads
\begin{equation}
E_\mathrm{L}\exp(-r^2/w^2_0)=\sum_{m=1}^\infty A_mE_{1m}=\sum_{m=1}^\infty A_m J_0\left(\frac{ u_m r}{R_{\mathrm{cap}}}\right)\ .
\label{eq:c4Econtinuity}
\end{equation}
The coefficient $A_m$ indicates the amplitude  of the EH$_{1m}$ mode, which can be determined using the orthogonality of Bessel functions,
\begin{equation}
A_m=2E_\mathrm{L}\frac{\int_0^1x\exp(-x^2 R_{\mathrm{cap}}^2/ w_0^2) J_0(u_mx)\mathrm{d}x}{J_1^2(u_m)}\ .
\label{eq:c4Am}
\end{equation}
The coupling coefficient, $C_m$, is  defined as the fraction of incident energy coupled to the EH$_{1m}$ mode, and can be written as
\begin{equation}
C_m=8\left(\frac{R_{\mathrm{cap}}}{w_0}\right)^2\frac{ \left[ \int_0^1x\exp(-x^2 R_{\mathrm{cap}}^2/ w_0^2)J_0(u_mx)\mathrm{d}x \right] ^2}{J_1^2(u_m)}\ .
\label{eq:c4Cm}
\end{equation}
$C_m$ thus depends on the mode order $m$, the capillary radius $R_{\mathrm{cap}}$, and the laser waist $w_0$. Figure \ref{fig:modecoupling} shows the dependence of the coupling coefficient of an incident Gaussian beam on the first four  eigenmodes, as a function of the  ratio of capillary radius to laser waist,  $R_{\mathrm{cap}}/w_0$. It shows that monomode coupling can be achieved when $R_{\mathrm{cap}}/w_0\simeq1.55$ (or $w_0/R_{\mathrm{cap}}\simeq0.65$). In this case, almost 98\% of the  incident laser energy is coupled to the fundamental EH$_{11}$ mode, and  only 1\% of laser energy is coupled to higher-order modes. The remaining 1\% of laser energy is the energy contained in the wings of the Gaussian function outside the capillary diameter, and it is lost into the material at the front surface of capillary tube.

For $1.2 \leq R_{\mathrm{cap}}/w_0 \leq 2$, more than 90\%  of the incident laser energy can be coupled to the first mode.
When $R_{\mathrm{cap}}/w_0\to0$, the laser size increases, and the fundamental mode is predominantly excited, although coupling efficiency decreases quickly. This is because the laser energy is outside the capillary diameter and hits the front surface of the capillary tube, as illustrated by the left-hand sketch in Fig.~\ref{fig:modecoupling}. As $R_{\mathrm{cap}}/w_0$ increases, the laser waist becomes small compared with the optimal waist for quasi-monomode coupling. In this case, the laser energy is nearly 100\% coupled into the capillary tube, as indicated by the sum. However,
the energy  coupled to the fundamental mode  decreases at the benefit  of the excitation of higher-order modes, resulting in undesired mode beating and severe laser attenuation during the propagation in the capillary tube.

\subsubsection{Coupling of an Airy beam}

Another case of interest is the coupling of a laser beam with an Airy-like distribution in the focal plane \cite{hecht02}.
The electric field amplitude for an Airy beam focused at the capillary entrance ($z=0$) is written as
\begin{equation}
E_\mathrm{A}=E_\mathrm{L}\frac{J_1(\nu_1r/r_0)}{r}\ ,
\end{equation}
where $\nu_1=3.8317$ is the first root of the equation $J_1(x)=0$, and $r_0$ is the radius corresponding to the first zero. The continuity of the electric field at the capillary entrance gives
\begin{equation}
E_\mathrm{L}\frac{J_1(\nu_1r/r_0)}{r}=\sum_{m=1}^\infty A_mE_{1m}=\sum_{m=1}^\infty A_m J_0\left(\frac{ u_m r}{R_{\mathrm{cap}}}\right)\ ,
\end{equation}
and the coupling coefficient, $C_m$,  can be calculated from
\begin{equation}
C_m=\frac{4}{J_1^2(u_m)}\left[\int_0^1J_1\left(\frac{\nu_1R_{\mathrm{cap}}}{r_0}x\right)J_0(u_mx)\mathrm{d}x\right]^2\ .
\label{eq:c4CmAiry}
\end{equation}

\begin{figure}
\centering
\includegraphics[width=9cm]{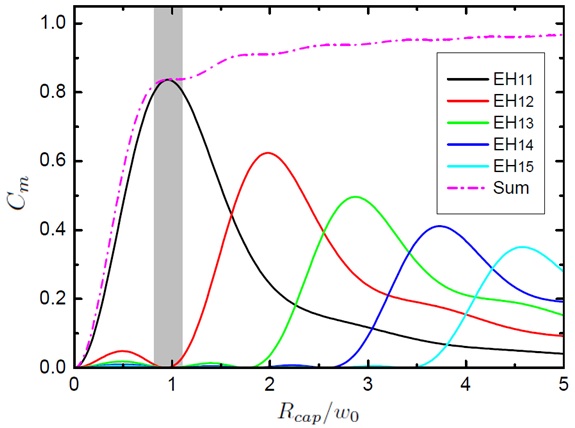}
\caption[Coupling efficiency of an Airy beam]{Coupling coefficient of an Airy beam to the first five eigenmodes as a function $R_{\mathrm{cap}}/r_0$. The grey area indicates the range of  $R_{\mathrm{cap}}/r_0$ where more than 80\% of the incident laser energy is coupled to the EH$_{11}$ mode.}
\label{fig:c4airy}
\end{figure}

The coupling coefficients for the first five eigenmodes are plotted  as functions of $R_{\mathrm{cap}}/r_0$ in Fig.~\ref{fig:c4airy}.
This shows that quasi-monomode guiding of an Airy beam can be achieved for $R_{\mathrm{cap}}/r_0\simeq1$. However, even in this case, only 83\% of the incident laser energy is coupled to the fundamental
mode, while the remaining 17\% energy contained by the laser distribution for $r>R_{\mathrm{cap}}$ hits the  capillary en\-trance wall. When $R_{\mathrm{cap}}/r_0$ increases, more laser energy can be coupled inside the capillary tube but
essentially to higher-order modes. It is thus  less efficient to use an Airy beam than a Gaussian beam in terms of laser coupling and monomode guiding. Another issue
is capillary breakdown. The front surface of the capillary tube is exposed to a greater energy for an Airy beam than for one with a Gaussian profile; hence, the peak laser intensity has to be lowered to avoid capillary damage at the entrance.

To summarize, quasi-monomode guiding can be selected by coupling the input laser energy to the fundamental EH$_{11}$ mode: 98\% of the energy of a Gaussian beam can be coupled to the fundamental mode for a waist size $w_0 = 0.645 R_{\mathrm{cap}}$. This mode  is preferable for laser wakefield acceleration as its group velocity is close to the velocity of light in free space, its damping factor is a minimum for a given capillary radius and wavelength, and the transverse electric field exhibits a variation similar to a Gaussian beam.

 \subsection{Experimental demonstration of laser guiding}

\begin{figure}
\begin{center}
\includegraphics[width=10cm]{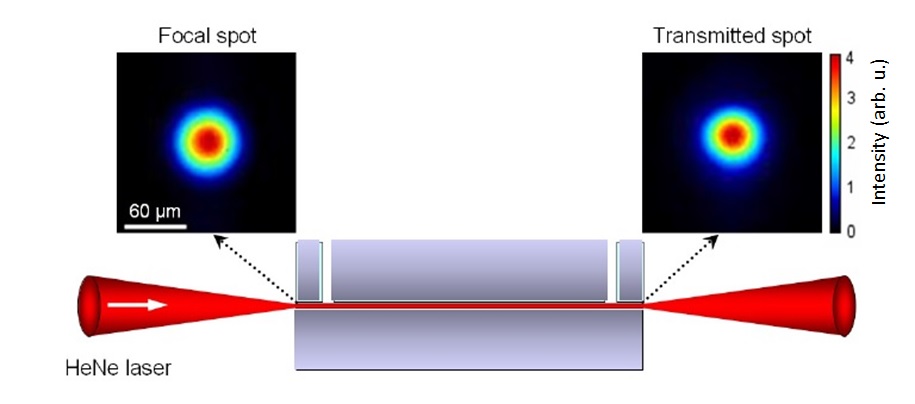}
\caption{Example of guiding of a low-intensity laser in the absence of plasma}
\label{fig:capguidhene}
\end{center}
\end{figure}

Guiding in capillary tubes at low intensity has been measured and corresponds to theoretical predictions in terms of coupling and transmission. Experimentally, the incident beam should be focused at the entrance plane of the capillary tube, as illustrated in Fig.~\ref{fig:capguidhene}. In this case, a HeNe laser ($\lambda_0 = 632\Unm$) was focused at the entrance of a capillary tube, of length $30\Umm$ and inner diameter $127\Uum$. The focal spot was measured to be Gaussian with a waist of about $43\Uum$, which corresponds to $w_0 = 0.68 R_{\mathrm{cap}}$, close to the matching condition for monomode guiding. Figure \ref{fig:capguidhene}
also shows the transmitted laser spot at the output of the capillary, which is symmetrical and exhibits the pattern of the fundamental mode.

In this case, a transmission of $90 \pm 2\%$ was obtained experimentally, and can be compared with the theoretical transmission. The transmission is
defined as the ratio of incident and transmitted laser energies at the capillary entrance and  exit, respectively.
The theoretical transmission of energy for the fundamental mode is given by $T_1 = C_1 \exp(-2L_{\mathrm{cap}}/L^l_1)$
The theoretical value of the coupling co\-efficient $C_1$ of a Gaussian beam for the  EH$_{11}$ mode is 97\% for  $w_0 = 0.68 R_{\mathrm{cap}} $. The attenuation length is calculated from Eq.~(\ref{eq:c4Lm})
as $L = 230\Ucm$ for $\lambda_0 = 632\Unm$ and $R_{\mathrm{cap}} = 63\Uum$. Thus the theoretical transmission is calculated as $T_1 = 94\%$, which is very close to the experimental value, suggesting that excellent alignment and beam quality were achieved in the experiment, leading to quasi-monomode guid\-ing. The sensitivity of the coupling and transmission to misalignments have been studied theoretically: predictions are in excellent agreement with experimental results and are discussed in Ref. \cite{veysman10}.

\begin{figure}
\begin{center}
\includegraphics[width=8cm]{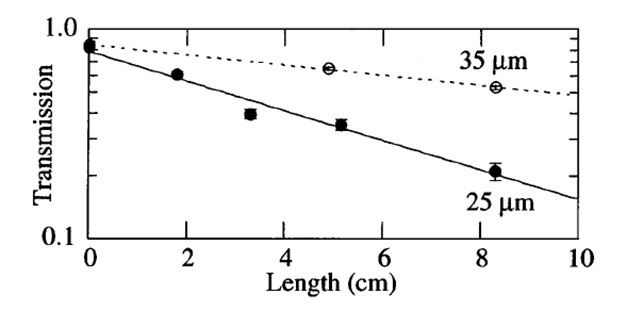}
\caption{Measured transmission as a function of propagation distance for two values of the capillary tube radius, $R_{\mathrm{cap}} = 35\Uum$ (open circles and dotted line), and $R_{\mathrm{cap}} = 25\Uum$ (filled circles and solid line).}
\label{fig:TL2Rcap}
\end{center}
\end{figure}

Monomode guiding has been demonstrated experimentally for laser intensities  of the order of $I_0 \simeq 10^{16}\UW/\UcmZ^2$  in a vacuum \cite{dorchies99} and the transmission has been measured  for different values of the capillary tube radius. In agreement with theoretical predictions, the damping length was found to increase with the cube of capillary radius, as illustrated in Fig.~\ref{fig:TL2Rcap}. The measured transmissions correspond to the predicted values for quasi-monomode guiding, which was thus measured in a vacuum over a distance of 100 Rayleigh lengths. Analytical predictions for coupling conditions and damping
length have been confirmed experimentally for tubes under vacuum.

\subsection{Measurement of plasma waves in capillary tubes}
The excitation of plasma waves over a length of up to $8\Ucm$ was, for the first time \cite{wojda2009}, demonstrated using laser guiding of intense laser pulses through hydrogen-filled glass capillary tubes. Laser guiding at input intensities up to $10^{18}\UW/\UcmZ^2$ was achieved with more than 90\% energy transmission in evacuated or hydrogen-filled gas tubes up to $8\Ucm$ long, with a radius of $R_{\mathrm{cap}}=50\Uum$. To investigate the linear and
moderately non-linear regime, the input intensity was
kept below $3 \times 10^{17}\UW/\UcmZ^2$, and the electron density was varied in the range $0.05$--$5 \times 10^{18}\Ucm^{-3}$. The
plasma wave amplitude was diagnosed by analyzing the spectrum of the transmitted laser radiation.
Laser pulses transmitted through gas-filled
capillary tubes exhibit broadened spectra. In the
range of parameters relevant to this experiment, spectral
modifications of the laser pulse driving the plasma wave, after propagating
in the plasma over a large distance, are mainly
related to changes in the index of refraction of the plasma
during the creation of the plasma wave. The front of the
laser pulse creates an increase in electron density, leading
to a blueshift at the front of the pulse, while the rear
of the pulse creates a decrease of electron density with
larger amplitude, and thus a redshift of the spectrum.

\begin{figure}
\begin{center}
\includegraphics[width=8cm]{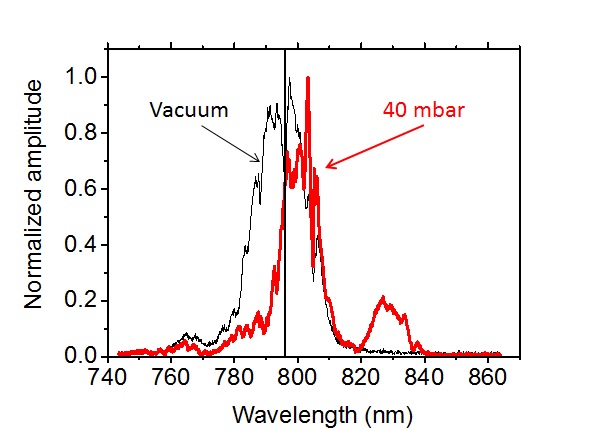}
\caption{Spectrum of the laser beam at the output of a capillary tube in vacuum (black solid curve, the vertical line indicates the centre of the spectrum in a vacuum through the capillary) and in $40\Umbar$ of hydrogen (red line).}
\label{fig:PWspectrum}
\end{center}
\end{figure}

The signature of plasma wave excitation is a redshift of the laser beam, \ie a shift of the spectrum towards longer wavelengths, as seen in Fig.~\ref{fig:PWspectrum}. The spectrum at the output of a $7\Ucm$ long capillary was measured in a vacuum and compared with the spectrum in the presence of $40\Umbar$ of hydrogen filling the tube.

\begin{figure}
\begin{center}
\includegraphics[width=7.5cm]{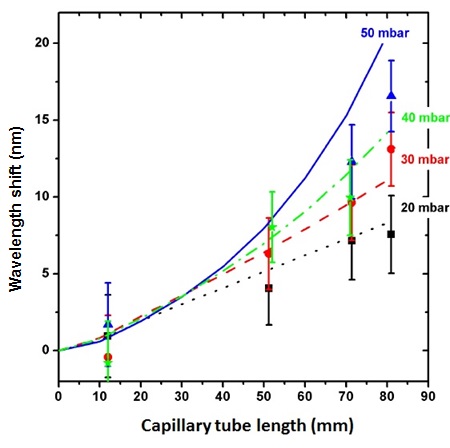}
\caption{Wavelength shifts as functions of the length of capillary filled with hydrogen at different pressure values, as determined experimentally (symbols with error bars) and through modelling (curves).}
\label{fig:shiftPlength}
\end{center}
\end{figure}

\begin{figure}
\begin{center}
\includegraphics[width=8cm]{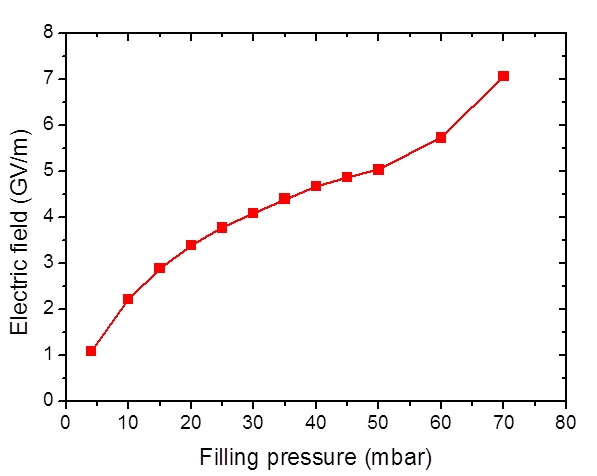}
\caption{Average amplitude of the plasma wave electric field measured at the output of a $8\Ucm$ long glass capillary tube as a function of filling pressure.}
\label{fig:PWelectricfield}
\end{center}
\end{figure}

The wavelength shift, $\Delta \lambda /\lambda$ can be shown  \cite{andreev05} to be directly related to the energy of the plasma wave with electric field, $E_\mathrm{p}$, excited in the
plasma  volume, $V$,
\begin{equation}
\frac{\Delta \lambda}{\lambda} \simeq \frac{1}{16 \pi \mathcal{E}_{\mathrm{out}}} \int _V E_\mathrm{p}^2 \mathrm{d}V\ ,
\end{equation}
where $\mathcal{E}_{\mathrm{out}} $ is the total energy of the transmitted laser pulse. For monomode propagation of a laser pulse with a Gaussian
time envelope, generating a wakefield in the weakly
non-linear regime, the wavelength shift can be expressed
analytically  \cite{wojda2009}. For small energy losses at the capillary wall, it is proportional to the peak laser intensity on the capillary axis, and to the
length of the capillary, and exhibits a resonant-like dependence
on gas pressure directly linked to plasma wave excitation. The dependence of the spectral redshift was measured as a function of filling pressure, capillary tube length, and incident laser energy, and was found to be  in excellent agreement with  results from modelling, as illustrated in Fig.~\ref{fig:shiftPlength}. A linear behaviour of the wavelength shift as a function of length is observed at $20\Umbar$. The fit of experimental data by simulation results
demonstrates that the plasma wave is excited over
a length as long as $8\Ucm$. As the pressure is increased,
the non-linear laser pulse evolution is amplified, with the
propagation length leading to a  plasma wave amplitude larger than the linear prediction.
The longitudinal accelerating field, inferred from a detailed analysis of laser wakefield dynamics in capillary tubes and numerical simulations  \cite{andreev10},  is in the range of 1--$10\UV[G]/\UmZ$ for an input laser intensity of the order of $I_0 \simeq 10^{17}\UW/\UcmZ^2$,
as shown in Fig.~\ref{fig:PWelectricfield}. The average product of gradient and length achieved in this experiment was thus of the order of $0.4\UV[G]$ at a pressure of $50\Umbar$; it could be increased to several gigavolts by increasing the length and diameter of the capillary tube with higher laser energy.

In conclusion, the outgoing spectra of driving laser pulses measured after propagation in gas-filled capillaries supported by relevant modelling provide detailed information on laser pulse dynamics and on the main characteristics of the accelerating fields excited in the wake of the laser pulses over the long distances necessary for efficient acceleration of electrons to high
energies.

\section{Conclusion and perspectives}

In summary, several features of laser-driven plasma wakefields linked to laser propagation in a plasma were introduced and discussed. Numerous results from  previous studies clearly indicate that reducing the plasma density  is the best method of increasing the energy of the accelerated electrons. At lower dens\-ities, such mechanisms as dephasing between the electron bunch and the plasma wave and the progressive depletion of the driving laser pulse have to be taken into account, as they limit the maximum attainable energy.
Increased electron energy of the order of a few gigaelectronvolts is currently achieved at lower plasma densities and higher laser peak powers, in non-linear regimes of laser plasma acceler\-ation, where extending the plasma length to the dephasing length is a challenge. Eventually, lowering the plasma density will lead to a regime where self-injection of plasma electrons in the plasma wave does not take place and injection of electrons from external sources has to be implemented.

Ongoing efforts of accelerator development tend to increase the energy of electrons, while im\-proving the electron bunch properties, of
which energy spread and transverse emittance are key para\-meters for
beam transport and focusing, as well as subsequent applications.
 Separating electron
injection and acceleration processes, and relying on a number of purely
accelerating stages  \cite{leemans09}, provide solutions to the issues of dephasing
and depletion, and make the acceleration process scalable to
higher energies, while preserving the bunch quality. Two crucial aspects of multistage laser-driven plasma acceleration are laser guiding over metre-scale distances \cite{paradkar2013} and control of the properties of the electron bunch for external injection into the plasma wave of the accelerating stage.

Future work will tackle the different issues of multistage laser-driven plasma acceleration. Laser guiding and increased laser energy are expected to produce electron bunches in the $10\UGeV$ range in one stage (see, for example, the BELLA project \cite{bella} in the USA or the CILEX APOLLON project \cite{cros2014} in France). Staging is the next milestone for the development of laser-driven plasma accelerators.
The APOLLON laser facility \cite{zou2015}, under construction in
France, will provide two beams in the 1--$10\UW[P]$ range, and a shielded large experimental area built to host multistage experiments \cite{cros2014}.
In those ex\-peri\-ments, the first stage (\ie the injector) will be used
to generate electrons in the 50--$200\UMeV$ range, which will then be
transported \cite{chance2014} and focused into the accelerator stage.
The quality of the accelerated bunch will thus result from the combined
intrinsic properties of the injector and the capability of the transport
line to accommodate them.

In the long term, the development of laser-driven accelerators will rely on improvements in  the performance of laser systems, in terms of beam quality, reliability, stability, and average power.
Plasma stages in the quasi-linear regime provide means to control transverse and longitudinal fields, and con\-sequently the dynamics of accelerated bunches. Electron or positron beams can be accelerated in this regime, and external injection schemes into metre-scale low-density plasma sources need to be de\-veloped.

\end{document}